\documentclass[prl,amsmath,a4paper,twocolumn]{revtex4}
\usepackage{graphicx}
\usepackage{amsfonts}
\usepackage{amsmath}
\usepackage{ltxtable}
\usepackage{CJK}
\usepackage{mdwmath}
\usepackage{color,soul}

\newcommand{\mN}{{\mathcal N}}

\newcommand{\mB}{{\mathcal B}}
\newcommand{\mC}{{\mathbb C}}
\newcommand{\mZ}{{\mathbb Z}}
\newcommand{\mV}{{\mathbb V}}
\newcommand{\mE}{{\mathbb E}}
\newcommand{\mK}{{\mathcal K}}
\newcommand{\mQ}{{\mathcal Q}}

\newcommand{\mI}{{\mathcal I}}
\newcommand{\mU}{{\mathcal U}}
\newcommand{\mmV}{{\mathcal V}}

\begin{document}

\title{Construction of State-independent Proofs for Quantum Contextuality}

\author{Weidong Tang$^{1}$
and Sixia Yu$^{2}$}
\affiliation{$^1$Key Laboratory of Quantum Information and Quantum Optoelectronic Devices, Shaanxi Province,
 and Department of Applied Physics
of Xi'an Jiaotong University, Xi'an 710049, P.R. China
\\
$^2$Hefei National Laboratory for Physical Sciences at
Microscale and Department of Modern Physics
of University of Science and Technology of China, Hefei 230026, P.R. China}

\begin{abstract}
Since the enlightening proofs of quantum contextuality first established by Kochen and Specker, and also by Bell, various simplified proofs have been constructed to exclude the non-contextual hidden variable theory of our nature at the microscopic scale. The conflict between the non-contextual hidden variable theory and quantum mechanics is commonly revealed by Kochen-Specker (KS) sets of yes-no tests, represented by projectors (or rays), via either logical contradictions or noncontextuality inequalities in a state-(in)dependent manner.  Here we first propose a systematic and programmable construction of a state-independent proof from a given set of nonspecific rays in $\mC^3$ according to their Gram matrix. This approach  brings us a greater convenience in the experimental arrangements. Besides, our proofs in $\mC^3$ can also be generalized to any higher dimensional systems by a  recursive method.

\end{abstract}

\maketitle

Determinism had once been considered as an axiom by mainstream physicists for hundred of years before the early $20$-th century. After that, difficulties arose when people tried to build a consistent theory (quantum theory) to interpret the seemingly random outcomes of measurements performed  on a microscopic system. To overcome the difficulties, one approach by considering the randomness as an underlying feature, namely, by retaining the notion of indeterminism, is known as Copenhagen interpretation of quantum mechanics, and another route, by trying to introduce an extra set of variables to make the theory deterministic, is known as a hidden variable theory\cite{Bell, CHSH}, e.g. Bohmian Mechanics.

However, people found that  some hidden variable theories with seemingly sound constraints cannot reproduce all the predictions of quantum  mechanics. For example, a local hidden variable theory claiming that the predetermined value of measuring an observable by an observer is  independent of the measurement arrangement for another spacelike separated observable, tries to extend the ``seemingly" probabilistic features of quantum  mechanics to a deterministic interpretation framework  by introducing some underlying inaccessible variables (hidden variables) consistent with local realism.
Moreover, by extending space-like separated observables to compatible (commutative) ones, local hidden variable theories can be generalized to non-contextual hidden variable theories (NCHVT) \cite{Bell2, KS, mermin1} which  claim that  the value from a measurement of an observable predetermined by some hidden variables is immune to the choice of any compatible observable that it might be measured alongside. Namely, this value is independent of contexts, or in other words, non-contextual, where a context is a set of mutually compatible observables.
The process of excluding NCHVT from quantum  mechanics, or proving NCHVT cannot reproduce all the quantum mechanical predictions, is usually referred to as a proof for the Kochen-Specker (KS) theorem \cite{KS}.

So far, various illuminating proofs for KS theorem have been proposed, some of which even have been carried out by experiment in various  physical systems \cite{Michler,Huang1,Kirchmair,Amselem,Bartosik,Moussa,Lapkiewicz,Amselem2,Zu,Vincenzo,XiangZhang,Huang2}.
To be specific, there are conventional KS proofs from sets of rays(rank-one projectors) by yes-no tests such as state-independent proofs from the 117-ray model to the 13-ray model \cite{KS,Peres,Bub,Conway,yu-oh} and state-dependent proofs from the family of $(2n+1)$-ray models\cite{KCBS,Liang} and  Clifton's 8-ray model \cite{Clifton}, KS inequalities\cite{KCBS,Simon,Larsson,cabello2,cabello3,TYO}, Peres-Mermin squares-like proofs\cite{mermin1,mermin1990,Perespla,Peresbook}, anomalous weak values' proof\cite{Pusey}, Hardy-like proof\cite{hardylike}, etc.   Cabello {\it et al}{\cite{NS condi}} found the necessary and sufficient condition to test whether a given set of rank-one projectors is  a state-independent contextual(SIC) set, i.e., can form a proof for the KS theorem in a state-independent manner.

A natural question arises:  how can we choose the relevant projectors to form such a SIC set?  More specifically, given several generally chosen rank-one projectors (with a pre-constraint but easy to be satisfied), how to economically add some auxiliary projectors to form a SIC set? Clearly, many well-known state-independent proofs for the KS theorem were special solutions to this question. After all, analytic construction for a proof usually requires some special symmetries of the set. Another way was by brute force search with the aid of a computer\cite{Peresbook,NS condi,AEG,PCs2}, but it was rather inefficient. So far, a method to construct a SIC set,  which is expected to be analytic, systematic, and easy to handle,  has not yet been developed.

\begin{figure}
\includegraphics[scale=0.7]{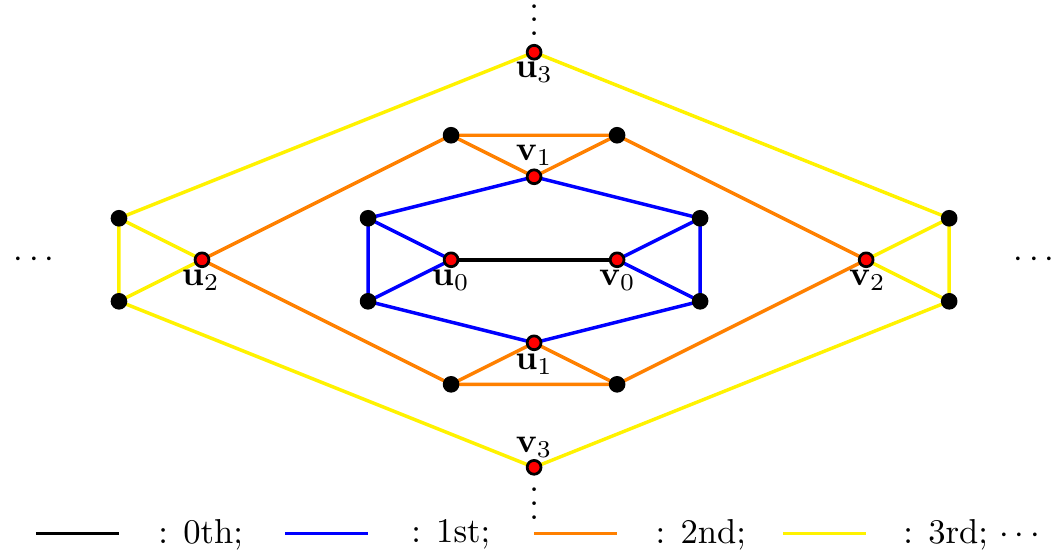} \caption{\label{n-order}The $(2+6n)$-ray model as a generalization of Clifton's 8-ray model. Each colored vertex represents a ray and each edge linking two vertices indicates their orthogonality.  The central $8$ rays --- $\textbf{u}_0,\textbf{u}_1,\textbf{v}_0,\textbf{v}_1$ and all the 4 unlabeled rays attached to $\textbf{u}_0$ and $\textbf{v}_0$, form the original Clifton's 8-ray model. The outer rays are used for the generalized ones.}
\end{figure}

In this paper, based on a general constraint between any pair of  nonparallel rays in $\mC^3$, which can be revealed by adding some auxiliary complete orthonormal bases to form a generalized Clifton's model(see FIG.\ref{n-order}), an  analytic method to construct a SIC proof for the KS theorem can be developed. By considering its universality, the proofs for the KS theorem can be adjusted flexibly according to the related experimental arrangements.  Moreover, we can recursively construct new proofs in $\mC^D$ ($D>3$) from our proofs in $\mC^3$.

For convenience, we will use the notation of a ray (normalized unless emphasized) to represent two different but one-to-one corresponding things: a complex vector in the projective Hilbert space, and a normalized rank-$1$ projectors on the ray.  That is, a projector is in the form of $P=|{\psi}\rangle\langle {\psi}|$, where the ray $|{\psi}\rangle=\alpha|0\rangle+\beta|1\rangle+\gamma|2\rangle$, which can also be denoted by $\textbf{r}=(\alpha,\beta,\gamma)$, with $\alpha,\beta,\gamma\in \mC$. Then the normalization, i.e., $\langle\psi|{\psi}\rangle=1$, indicates that $\textbf{r}^{\ast}\textbf{r}=1$. By contrast, the orthogonality, namely, $\langle\psi_1|{\psi}_2\rangle=0$, implies that $\textbf{r}_1^{\ast}\textbf{r}_2=\textbf{r}_2^{\ast}\textbf{r}_1=0$, and vice versa.
Notice that unlike common proofs for KS theorem from the real projective Hilbert spaces, here our proofs are extended to complex ones.

The KS value assignment rules guide our search for sets of rays that reveal quantum
contextuality. These rules are as follows:  (I) two mutually orthogonal rays cannot be assigned to value $1$ simultaneously, and (II) for the rays forming a complete orthonormal basis, one and only one of them must be assigned to value $1$.

First, we extend the Clifton's 8-ray state-dependent proof\cite{Clifton} for KS theorem to a complex Hilbert space. The orthogonality relationship of the $8$ rays is shown in FIG.\ref{n-order} as a subgraph consisting of 8 vertices connecting with blue or black edges.
The modulus for the inner product of two rays $\textbf{u}_1$ and $\textbf{v}_1$, denoted by $|\textbf{u}_1^*\cdot \textbf{v}_1|$, should satisfy $|\textbf{u}_1^*\cdot \textbf{v}_1|\leq\delta_1=\frac{1}{3}$ (See Supplementary Material). Considering the KS value assignments to $\textbf{u}_1$ and $\textbf{v}_1$,  assigning value $1$ to them together will lead to a contradiction that two orthogonal rays $\textbf{u}_0$ and $\textbf{v}_0$  must be assigned to value $1$ simultaneously.

Likewise, analogous  value assignment contradictions  can be generalized to the other labeled rays in FIG.\ref{n-order}, e.g., assigning value $1$ to the rays $\textbf{u}_2$ and $\textbf{v}_2$ simultaneously will give rise to a  contradiction that the rays $\textbf{u}_1$ and $\textbf{v}_1$ should be assigned to value $1$. Then we have the following lemma.

{\it Lemma ---} Two rays  $|\psi\rangle$ and $|\phi\rangle$ with their inner product satisfying
\begin{align}\label{Dn}
|\langle\psi|\phi\rangle|\le{\frac{n}{n+2}}:=  \delta_n,
\end{align}
where $n$ is an non-negative integer, cannot be assigned to value 1 simultaneously according to KS value assignment rules after introducing $2n$ suitable complete orthonormal bases.

The case for $n=0$ is trivial. The proof for other cases presented in Supplementary Material(SM) is constructive, i.e., $2n$ complete orthonormal bases
can be found explicitly. Their orthogonality graph is shown in FIG.\ref{n-order}. From this orthogonality graph, if we denote by $P_{\psi}=|\psi\rangle\langle\psi|=\textbf{u}_n$ and $P_{\phi}=|\phi\rangle\langle\phi|=\textbf{v}_n$ ,   then the following state-dependent non-contextual inequality can be derived:
\begin{equation}\label{SINCEQcell}
    \langle P_\psi\rangle+\langle P_\phi\rangle+\langle C(\psi,\phi)\rangle\leq 2n+1
\end{equation}
with maximal quantum violation $2n+1+|\langle\psi|\phi\rangle|$ (see SM),
where
\begin{eqnarray}\label{c}
C(\psi,\phi)=\sum_{i\in V}P_i-\sum_{i<j,(i,j)\in E}P_iP_j-P_{\psi}-P_{\phi}.
\end{eqnarray}
Here in the expansion of $C(\psi,\phi)$, $V$ and $E$ denote the vertex set and the edge set for the orthogonality graph respectively, and $P_i$ stands for the $i$-th ray. We call the constraint
of the value assignments to the rays $|\psi\rangle$ and $|\phi\rangle$ in the Lemma the {\it $n$-th order KS value assignment constraint}.

Two remarks are in order.  First, our  recursive method to derive the KS value assignment constraints  is considerably cost-effective, i.e.,  at the price of only several extra complete orthonormal bases. Second, it is clear that the lower order KS value assignment constraint can be covered by a higher order constraint but is more more economical (with less auxiliary complete orthonormal bases to reveal that) within the same range of the inner product for two rays.

To a set of rays $\{|\psi_j\rangle\}_{j\in I}$ we associate with an integer $M$, given integer $n$. Let $G$ be a graph with its adjacency matrix $\Gamma$ defined by
$\Gamma_{ij}=1$ if $|\langle\psi_i|{\psi}_j\rangle|>\delta_n$ $(i\neq j, i,j\in I)$ and $\Gamma_{ij}=0$ otherwise, i.e., this graph only depends on the Gram matrix of the set of rays $\{|\psi_j\rangle\}_{j\in I}$. Then we denote by $M$ the vertex number of the largest clique $C_{\max}$, i.e., $M=|C_{\max}|$, where a {\it clique} $C$ of the graph $G$ is defined by $C\subseteq I$ such that $\Gamma_{kl}=1$ for all $k,l\in C$ $(k\neq l)$. Moreover we denote the minimal eigenvalue of $\sum_{j\in I}P_j$ by $\lambda_{\min}$, where $P_j=|\psi_j\rangle\langle\psi_j|$. If $M< \lambda_{\min}$ then we call $\{P_j\}_{j\in I}$ an $n$-th order {\it fundamental KS ray set}(FKRS).

{\it Theorem 1. ---} Any $n$-th order FKRS can be completed by adding a finite number of orthonormal bases to give a state-independent proof for the KS theorem.

{\it Proof}.--- Specifically we shall prove that given an $n$-th order FKRS $\{P_j\}_{j\in I}$, $\{P_j\}_{j\in I}\cup\{Q_k\}_{k\in J}$ gives a  state-independent proof for the KS theorem, where  $\{Q_k\}_{k\in J}$ is the set of rays from all the auxiliary complete orthonormal bases (no rays overlapped unless specified) given as following. For each $m=1,2,\ldots, n$ and $(r,s)\in \alpha_m$, where
$$\alpha_m=\{(i,j)|i,j\in I,i>j, \delta_{m-1}< |\langle \psi_i|\psi_j\rangle|\le \delta_m\}$$
we introduce $2m$ complete orthonormal bases as in Lemma    such that these  bases together with $P_r$ and $P_s$ form a $(2+6m)$-ray model corresponding to the orthogonality graph in FIG.\ref{n-order} and there are in total $|J|=3R$ rays where $R=\sum_{m=1}^n2 m|\alpha_m|$.

Conventionally, KS theorem is proved by looking for a logical contradiction of the KS value assignment --- or in other words, the KS value assignment does not exist --- to a specially arranged set of rays. Here we use a similar argument from the Yu-Oh model\cite{yu-oh}.
Though all possible KS value assignments to our sets of rays might exist, the predictions of quantum  mechanics cannot be produced.

From the construction of the graph $G$  and the definition of its largest clique $C_{\max}$, we can infer that at most $M$ rays in the FKRS $\{P_j\}_{j\in I}$ can be simultaneously assigned to value $1$. Because if $|\langle\psi_i|{\psi}_j\rangle|\leq\delta_n$ $(i,j\in I)$, then $P_i$ and $P_j$ can not be assigned to $1$ simultaneously. This can be seen with the help of
the auxiliary rays from $\{Q_k\}_{k\in J}$ by the KS value assignment rules.

Next denoting by $P_j^{\lambda}\in\{0,1\}$ the value assigned to $P_j$ for a given $\lambda$, apparently we have $\sum_{j\in I}P_j^{\lambda}\leq M$. Moreover, any NCHV model which admits a KS value assignment should satisfy the following inequality
\begin{equation}\label{NCHV ineq}
  \sum_{j\in I}\langle P_j\rangle_c\equiv\sum_{j\in I}\int d\lambda\rho_{\lambda}P_j^{\lambda}\leq M.
\end{equation}
However, quantum  mechanics  predicts
$\lambda_{\min}\leq\sum_{j\in I}\langle P_j\rangle_q$.
And from the definition of the $n$-th order FKRS, we have $M<\lambda_{\min}$. That is, any NCHV model obeying the KS value assignment rules cannot reproduce all the predictions of quantum  mechanics by any state. Generally, the auxiliary set $\{Q_k\}_{k\in J}$ cannot give a state-independent proof for KS theorem itself (this condition can be easily satisfied since usually there is  a range choice for the directions of the  complete orthonormal bases added, and they can always be adjusted to some nonspecific directions).

If we do not invoke the KS value assignment rules then we can construct the following state-independent non-contextual inequality
\begin{align}\label{SINCEQ-2}
    \sum_{i\in I}\langle P_i\rangle+\sum_{(i,j)\in I, i>j}C(P_i,P_j)\le M+R.
\end{align}
where $R$ is the total number of complete orthonormal bases added and $C(P_i,P_j)$ is defined in Eq.(\ref{c}) which vanishes if $|\langle\psi_i|\psi_j\rangle|>\delta_n$. Details for the proof can be found in SM. Obviously, Eq.(\ref{SINCEQ-2}) can be violated by quantum  mechanics, as the left hand side is no less than $R+\lambda_{\min}>R+M$. Hence all the rays in $\{P_j\}_{j\in I}\cup\{Q_k\}_{k\in J}$, provide a state-independent proof for the KS theorem. \hfill$\sharp$

In fact, this theorem provides us a general method to look for a proof for the KS theorem  in a state-independent fashion. After all, an FKRS can easily be generated, whether by adding/deleting several rays or adjusting the directions of some existing ones from a non-FKRS. The choice of the auxiliary complete orthonormal bases should also be careful, as for some special cases (e.g., in Ref.\cite{yu-oh}) part of the rays from one complete orthonormal basis may overlap with the rays from other bases and may simplify the proof.

Besides, by converting the KS value assignment of any given set of rays to an old problem in graph theory, i.e., looking for the largest clique for a given graph,  one may clearly see how an FKRS plays a central role in a proof for the KS theorem and can easily catch essence of  the conflict between NCHVT and quantum  mechanics.

{\it Example 1}.--- $4$ rays $P_1,P_2,P_3$ and $P_4$ orienting to $4$ vertices of a regular tetrahedron respectively form a first order FKRS.

Notice that $P_i=|\psi_i\rangle\langle\psi_i|$, and $|\langle\psi_i|{\psi}_j\rangle|=\frac{1}{3}$ for all $i\neq j$ $(i,j=1,2,3,4)$. As  $\delta_1=\min{\delta_n}=\frac{1}{3}$, all the entries of the adjacency matrix are 0. Hence any single vertex can be regarded as the largest clique, i.e., $M=1$.
By contrast, we have $\lambda_{\max}=\lambda_{\min}=\frac{4}{3}>M$. Therefore, these $4$ rays can form a first order FKRS and can be used for providing a state-independent proof for KS theorem according to Theorem 1.  Actually, this FKRS can give rise to  the 13 rays' state-independent proof by Yu and Oh\cite{yu-oh} by adjusting the auxiliary complete orthonormal bases to let some  rays from them overlap with each other by Theorem 1. This can also help us to understand the minimality for the ray number in Yu-Oh model from a more essential point as after deleting any one from the four rays, the remained can no longer form an FKRS.

In the following we give another example first appearing in Ref.\cite{SIC21}.

{\it Example 2. --- } A set of 9 rays (un-normalized)
\begin{equation}
   \begin{array}{ccc}
      \{(0,1,-1), &  (1,0,-1),& (1,-1,0), \cr
      (0,1,-\omega), & (1,0,-\omega), & (1,-\omega,0), \cr
    (0,1,-\omega^2), & (1,0,-\omega^2), & (1,-\omega^2,0)\},
    \end{array}
\end{equation}
is a second order FKRS, where $\omega=e^{i\frac{2\pi}{3}}$. This can be seen from the fact that after normalization the inner product for each pair of rays is $\frac{1}{2}$.

Although Cabello {\it et al} presented the necessary and sufficient condition to identify whether a given proof for the KS theorem is state-independent or not in Ref.\cite{NS condi}, they did not provide a method to build it.  To settle this problem, a systematic approach can be developed by Theorem 1, as a supplementary of Ref.\cite{NS condi}, which seems quite general and effective.

Further, as for any given set of rays $\{|\psi_j\rangle\}_{j\in I}$, there exists a natural number $N$ such that for all $i\neq j, i,j\in I$, we have $\delta_{N-1}<\max |\langle\psi_i|\psi_j\rangle|\leq\delta_{N}$.
From Theorem 1, by using the $N$-th order KS value assignment constraint, namely, in the case of $n=N$, we can always built a graph $G$ whose adjacency matrix $\Gamma=0$. Thus, we get the following corollary.

{\it  Corollary 1.---} Given a set of $n=|I|$ rays $\{|\psi_j\rangle\}_{j\in I}$ satisfying $\sum_{i\in I}|\psi_i\rangle\langle\psi_i|>1$,  we can always construct a state-independent proof for KS theorem by adding to this set at most $n(n-1)N$  complete orthonormal bases,  where  $$N=\left\lceil\frac{2|\delta|_{\max}}{1-|\delta|_{\max}}\right\rceil,\quad |\delta|_{\max}\equiv\max_{i\neq j (i,j\in I)}|\langle\psi_i|\psi_j\rangle|.$$

As $\frac{2|\delta|_{\max}}{1-|\delta|_{\max}}\leq N<\frac{2|\delta|_{\max}}{1-|\delta|_{\max}}+1$,  we have $\frac{N-1}{N+1}=\delta_{N-1}<|\delta|_{\max}\leq\delta_N=\frac{N}{N+2}$. And the proof is straightforward.

For any dimension $D>3$, there is a recursive method to construct a proof for the KS theorem  based on a given proof of dimension $d$\cite{KP,AEG,HD}. Given a set of rays $S_{d}$ that can give us a state-independent proof in dimension $d$,  for any dimension $D=d+m,~1\leq m\leq d$, we can introduce    $S^+_{d+m}=\{(\textbf{v},\textbf{0})|\textbf{v}\in S_d,\textbf{0}\in\mC^{m}\}\cup\{(\textbf{0},\textbf{a})|\textbf{0}\in \mC^d,\textbf{a}\in B_m\}$ and $S^-_{d+m}=\{(\textbf{0},\textbf{v})|\textbf{v}\in S_d,\textbf{0}\in\mC^{m}\}\cup\{(\textbf{a},\textbf{0})|\textbf{0}\in \mC^d,\textbf{a}\in B_m\}$, where $B_m$ is a complete orthonormal basis in $\mC^m$. Then $S^+_{d+m}\cup S^-_{d+m}$ can provide us a state-independent proof for the KS theorem in $\mC^D$.  As a proof in $\mC^3$ can be easily constructed by our method, then by recursion, a proof in any higher dimension can also be constructed.
From this point of view, construction of a state-independent proof for the KS theorem in $\mC^3$ is quite significant in the issue of quantum contextuality.

In summary, by introducing a $(2+6n)$-ray model to construct a new KS value assignment constraint, we have presented a general analytic method for constructing state-independent proofs for the KS theorem and get a practical state-independent non-contextual inequality.  Our  methods can find far more families of proofs than any existing constructions.  Not only do they have advantages such as being analytic, flexible, they can also enable us to get various classical bounds for the sum of the rays in a projector set by different orders of KS value assignment constraints, which may facilitate more detailed studies about the contextuality, e.g., the level of the quantum violation for the state-independent inequality from few-ray proofs to multi-ray proofs etc. Moreover, for those enthusiasts preferring to get new proofs for KS theorem by computer searching, our methods can  help them to check their outcomes with considerably high efficiency.

\begin{acknowledgements}
This work is supported by the NNSF of China (Grant No. 11405120) and the Fundamental Research Funds for the Central Universities.
\end{acknowledgements}

\newpage

\section{Appendix A. Proof of Lemma }

\begin{figure}[h]
\includegraphics[scale=1]{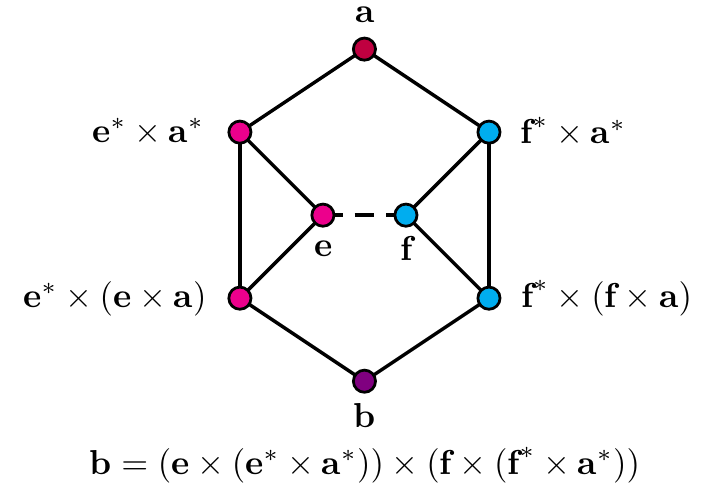} \caption{\label{CCli} The modified Clifton's 8-ray model. The rays in the complex Hilbert space $\mC^3$, are represented by the colored dots,  and any solid line linking two rays indicates the orthogonality for them. }
\end{figure}

(i) First, a state for a qutrit can be written as $|\psi\rangle=\alpha|0\rangle+\beta|1\rangle+\gamma|2\rangle$, and can be represented by a complex ray $\textbf{r}=\left(
                 \begin{array}{c}
                  \alpha \\
                  \beta \\
                   \gamma \\
                 \end{array}
               \right)$. Thus, we have (1) for any $|\psi\rangle$, $\langle\psi|\psi\rangle=1\Leftrightarrow \textbf{r}^{\ast}\cdot\textbf{r}=|\textbf{r}|^2=1$; (2) for any $|\psi_1\rangle,|\psi_2\rangle$, $\langle\psi_1|\psi_2\rangle=0\Leftrightarrow \textbf{r}_1^{\ast}\cdot\textbf{r}_2=\textbf{r}_2^{\ast}\cdot\textbf{r} _1=0$ (orthogonality).

(ii) The orthogonality of two  rays in a complex Hilbert space can be defined by the scalar product as referred above. Therefore, if a ray $\textbf{c}$ satisfies $\textbf{c}^{\ast}\cdot \textbf{e}=\textbf{c}^{\ast}\cdot \textbf{a}=0$ $(\textbf{e}\neq\textbf{a})$, then we have $\textbf{c}\propto\textbf{e}^{\ast}\times\textbf{a}^{\ast}$, and vice versa.
\subsection{From orthogonality graph to inequality }

Next we shall prove that if for two given rays $|\psi\rangle $ and $|\phi\rangle$  there exists two rays $|e\rangle $ and $|f\rangle$ such that the orthogonality graph FIG.1 is possible then
$$|\langle \psi|\phi\rangle|\le \frac {1+|\delta|}{3-|\delta|},\quad \delta=\langle e|f\rangle. $$

(iii) Let $\textbf{g}=\textbf{e}^{\ast}\times\textbf{f}^{\ast}$, and from the fact that $\textbf{e}^{\ast}\cdot\textbf{f}=\delta$ with $|\delta|<1$, we can get $|\textbf{g}|^2=|\textbf{e}^{\ast}\times\textbf{f}^{\ast}|^2=1-|\delta|^2>0$.

(iv) Let the ray $\textbf{a}$ correspond to $|\psi\rangle$ and it can be written as $$\textbf{a}=x\textbf{e}+y\textbf{f}+z\frac{\textbf{e}^{\ast}\times\textbf{f}^{\ast}}{\sqrt{1-|\delta|^2}}.$$ Note that $\textbf{e}$ is not orthogonal to $\textbf{f}$ in general.

(v) We have $a_e=\textbf{e}^{\ast}\cdot\textbf{a}=x+\delta y$ and $a_f=\textbf{f}^{\ast}\cdot\textbf{a}=x\delta^{\ast}+y$. To normalize $\textbf{a}$, we should ensure that $1=|\textbf{a}|^2=|x|^2+|y|^2+|z|^2+xy^{\ast}\delta^{\ast}+x^{\ast}y\delta$.

(vi) Let us introduce $I=|a_e|^2+|a_f|^2-a_e^{\ast}a_f\delta-a_ea_f^{\ast}\delta^{\ast}$. Then
\begin{align*}\label{I}
   I=&(|x|^2+|y|^2)(1+|\delta|^2)+2(xy^{\ast}\delta^{\ast}+x^{\ast}y\delta)\\
   -&[(|x|^2+|y|^2)|\delta|^2+x^{\ast}y\delta+xy^{\ast}\delta^{\ast}|\delta|^2]\\
   -&[(|x|^2+|y|^2)|\delta|^2+xy^{\ast}\delta^{\ast}+x^{\ast}y\delta|\delta|^2]\\
   =&(|x|^2+|y|^2+xy^{\ast}\delta^{\ast}+x^{\ast}y\delta)(1-|\delta|^2)\\
   =&(1-|z|^2)(1-|\delta|^2).
\end{align*}

(vii)As is known, if $r_1, r_2, r\in\mC$, then $|r_1|^2+|r_2|^2\geq2|r_1|\cdot|r_2|$ and $r+r^{\ast}\leq2|r|$. Thus, by the definition, we have $I\geq|a_e|^2+|a_f|^2-2|\delta|\cdot|a_e|\cdot|a_f|\geq2|a_e|\cdot|a_f|\cdot(1-|\delta|)$. Combining with (vi), we can get
\begin{equation*}\label{ineq1}
   2|a_e|\cdot|a_f|\leq(1+|\delta|)\cdot(1-|z|^2).
\end{equation*}

(viii) Next, we can write $\textbf{b}$, which corresponds to $|\phi\rangle$, as an unnormalized form
\begin{align*}\label{b}
    \textbf{b}=&(\textbf{e}\times(\textbf{e}^{\ast}\times \textbf{a}^{\ast}))\times(\textbf{f}\times(\textbf{f}^{\ast}\times \textbf{a}^{\ast}))\\
    =&(\textbf{a}^{\ast}-a^{\ast}_e\textbf{e}^{\ast})\times(\textbf{a}^{\ast}-a^{\ast}_f\textbf{f}^{\ast})\\
    =&a^{\ast}_ea^{\ast}_f(\textbf{e}^{\ast}\times\textbf{f}^{\ast})-a_e^{\ast}(\textbf{e}^{\ast}\times \textbf{a}^{\ast})-a_f^{\ast}(\textbf{a}^{\ast}\times \textbf{f}^{\ast})
\end{align*}
by noticing that $\textbf{A}\times(\textbf{B}\times \textbf{C})=\textbf{B}(\textbf{A}\cdot \textbf{C})-\textbf{C}(\textbf{A}\cdot \textbf{B})$.

(ix) Then, $\textbf{a}^{\ast}\cdot \textbf{b}=a^{\ast}_ea^{\ast}_fa^{\ast}_g$, where $a^{\ast}_g=\textbf{a}^{\ast}\cdot\textbf{g}=\textbf{a}^{\ast}\cdot(\textbf{e}^{\ast}\times\textbf{f}^{\ast})$. From (iv), we can see that $|a_g^{\ast}|^2=|z|^2(1-|\delta|^2)$.

(x) Analogous to (iii), $\textbf{e}^{\ast}\cdot \textbf{a}=a_e\Rightarrow|\textbf{e}^{\ast}\times\textbf{a}^{\ast}|^2=1-|a_e|^2$, and $\textbf{f}^{\ast}\cdot \textbf{a}=a_f\Rightarrow|\textbf{f}^{\ast}\times\textbf{a}^{\ast}|^2=1-|a_f|^2$. Then
\begin{align*}
    &|\textbf{b}|^2=\textbf{b}^{\ast}\cdot\textbf{b}\\
    =&|a_ea_f|^2\cdot|\textbf{e}^{\ast}\times\textbf{f}^{\ast}|^2+|a_e|^2(1-|a_e|^2)+|a_f|^2(1-|a_f|^2)\\
    &+a_e^{\ast}a_f(\textbf{e}^{\ast}\times\textbf{a}^{\ast})\cdot(\textbf{a}\times\textbf{f})
    +a_ea_f^{\ast}(\textbf{e}\times\textbf{a})\cdot(\textbf{a}^{\ast}\times\textbf{f}^{\ast})\\
    &-|a_e|^2a_f^{\ast}(\textbf{e}^{\ast}\times\textbf{f}^{\ast})\cdot(\textbf{e}\times\textbf{a})
    -|a_e|^2a_f(\textbf{e}\times\textbf{f})\cdot(\textbf{e}^{\ast}\times\textbf{a}^{\ast})\\
    &-|a_f|^2a_e^{\ast}(\textbf{e}^{\ast}\times\textbf{f}^{\ast})\cdot(\textbf{a}\times\textbf{f})
    -|a_f|^2a_e(\textbf{e}\times\textbf{f})\cdot(\textbf{a}^{\ast}\times\textbf{f}^{\ast}).
\end{align*}
Using $(\textbf{A}\times \textbf{B})\cdot(\textbf{C}\times \textbf{D})=\textbf{A}\cdot[\textbf{C}(\textbf{B}\cdot \textbf{D})-\textbf{D}(\textbf{B}\cdot \textbf{C})]$ and (vi), we obtain
\begin{equation*}
    |\textbf{b}|^2=|a_ea_f|^2(1-|\delta|^2)+I(1-|a_e|^2-|a_f|^2).
\end{equation*}

(xi) From (ix), we notice that $|\textbf{a}^{\ast}\cdot \textbf{b}|^2=|a_ea_f|^2|z|^2(1-|\delta|^2)$. We have
\begin{equation*}
    \frac{|\textbf{b}|^2}{|\textbf{a}^{\ast}\cdot \textbf{b}|^2}=\frac{1}{|z|^2}+\frac{1-|z|^2}{|z|^2}\cdot\frac{1-|a_e|^2-|a_f|^2}{|a_e|^2|a_f|^2}.
\end{equation*}
Combining with (vii), we get
\begin{widetext}
\begin{align*}
     \frac{|\textbf{b}|^2}{|\textbf{a}^{\ast}\cdot \textbf{b}|^2}\geq
     &\frac{1}{|z|^2}+\frac{1-|z|^2}{|z|^2}\cdot\frac{1-I-2|a_ea_f|\cdot|\delta|}{|a_e|^2|a_f|^2}\\
     \geq&\frac{1}{|z|^2}+\frac{1-|z|^2}{|z|^2}\cdot\frac{1-(1-|z|^2)(1-|\delta|^2)-(1-|z|^2)(|\delta|+|\delta|^2)}{|a_e|^2|a_f|^2}\\
     \geq&\frac{1}{|z|^2}+\frac{4}{|z|^2}\cdot\frac{1-(1+|\delta|)(1-|z|^2)}{(1+|\delta|)^2(1-|z|^2)}
     =\frac{1}{(1+|\delta|)^2}\left(\frac{(1-|\delta|)^2}{|z|^2}+\frac{4}{1-|z|^2}\right)\\
     \equiv&\frac{1}{(1+|\delta|)^2}h(\chi),
  \end{align*}
\end{widetext}
where $h(\chi)=\frac{\rho}{\chi}+\frac{\tau}{1-\chi}$, $\chi=|z|^2\in(0,1)$, $\rho=(1-|\delta|)^2$ and $\tau=4$.
As $\frac{\partial}{\partial\chi}h(\chi)=-\frac{\rho}{\chi^2}+\frac{\tau}{(1-\chi)^2}$  and $\frac{\partial^2}{\partial\chi^2}h(\chi)=\frac{2\rho}{\chi^3}+\frac{2\tau}{(1-\chi)^3}>0$,
then $h(\chi)$ can reach its minimum when $\frac{\partial}{\partial\chi}h(\chi)=0$. It is not difficult for us to see that if and only if $|z|=\sqrt{\frac{1-|\delta|}{3-|\delta|}}$, $h(\chi)_{\min}=(3-|\delta|)^2$. Therefore
 \begin{align*}
\frac{|\textbf{b}|^2}{|\textbf{a}^{\ast}\cdot \textbf{b}|^2} \geq\left(\frac{3-|\delta|}{1+|\delta|}\right)^2,
\end{align*}
with equality if and only if $|z|=\sqrt{\frac{1-|\delta|}{3-|\delta|}}$.
Finally, we obtain $$|\langle\psi|\phi\rangle|=\frac{|\textbf{a}^{\ast}\cdot\textbf{b}|}{|\textbf{b}|}\leq\frac{1+|\delta|}{3-|\delta|}.$$

\subsection{From inequality to orthogonality graph}

(xii) Let us define a sequence $(\delta_0,\delta_1,\delta_2,...,\delta_n,...)$ by starting with $\delta_0=|\delta|=0$, i.e., $\textbf{e}$ is orthogonal to $\textbf{f}$, then the orthogonality graph FIG.\ref{CCli} corresponds to the Clifton's 8-ray model and we have $|\langle\psi|\phi\rangle|\le 1/3=\delta_1$. Recursively, if $|\langle e|f\rangle|=\delta_n$ then
\begin{equation*}
  \max|\langle\psi|\phi\rangle|=\delta_{n+1}=\frac{1+\delta_n}{3-\delta_n}.
\end{equation*}
which gives rise to
\begin{equation*}
  \delta_n=\frac{n}{n+2}.
\end{equation*}
For rays in the real Hilbert spaces, let $\theta_n$ is the angle between $\textbf{a}$ and $\textbf{b}$,
and clearly we have $|\cos\theta_n|=\delta_n$.

On the other hand, we shall prove that given two rays $|\psi\rangle $ and $|\phi\rangle$, if
$$|\langle \psi|\phi\rangle|\le\delta_n=\frac {1+\delta_{n-1}}{3-\delta_{n-1}}$$
for some $n$, then we can always find two rays $|e\rangle $ and $|f\rangle$ to make the orthogonality graph FIG.1 possible.

(xiii)The normalized rays corresponding to $|\psi\rangle$ and $|\phi\rangle$ are denoted by $\bf a$ and $\bf b$ respectively, then we define $\textbf{e}=\textbf{e}_+$ and $\textbf{f}=\textbf{e}_-$ with
\begin{eqnarray}\label{rc}
{\bf e_\pm}= \frac{\sin u}2 ({\bf a}-{\bf b} e^{iv})\pm \frac{e^{i\theta}\cos u}{1+c}({\bf a^*}\times {\bf b^*})+\nonumber \\
\frac{\sqrt{4c+(1-c)^2\sin^2 u}}{2(1+c)}({\bf a}+{\bf b} e^{iv})
\end{eqnarray}
with $\langle\psi|\phi\rangle=c e^{-iv}$, $c=|\langle\psi|\phi\rangle|\le \delta_n$, and $u,\theta$ being arbitrary. It is clear that the orthogonality graph FIG.\ref{CCli} holds, which is ensured by ${\bf(e_\pm\times a)\cdot(e^*_\pm\times b^*)}=0$, and $$|{\bf e^*\cdot f}|=1-\frac{2(1-c)\cos^2u}{1+c}.$$
It follows that $1\ge |{\bf e^*\cdot f}|\ge (3c-1)/(1+c)$ and since $\delta_{n-1}\geq (3c-1)/(1+c)$ from $c\le \delta_n$ we can find a suitable $u$ such that $|{\bf e^*\cdot f}|=\delta_{n-1}$. In fact we can choose
$$\cos u=\sqrt{\frac{(1-\delta_{n-1})(1+c)}{2(1-c)}}.$$
In the case of $c=\delta_n$ we can simply choose $u=0$.

If we denote ${\bf a} ={\bf e}_{n+1,+}$, ${\bf b} ={\bf e}_{n+1,-}$, and ${\bf e}_\pm ={\bf e}_{n\pm}$ then the  Eq.(\ref{rc}) defines recursively $n$ pairs of rays  ${\bf e}_{k\pm}$  satisfying $|{\bf e}^*_{k+}\cdot {\bf e}_{k-}|=\delta_{k-1}$ with $k=1,2,\ldots, n$. The orthogonality relations in FIG.\ref{CCli} in the main text is satisfyied and those $2n$ complete orthonormal bases can be constructed explicitly
$$\{\Delta_0^{k\pm}={\bf e}_{k\pm},\quad \Delta_\sigma^{k\pm}\propto {\bf e}_{k+1,\sigma}^*\times {\bf e}_{k,\pm}\ (\sigma=\pm)\}.$$

\subsection{Summary}

The above calculations give us a useful recursive relation between two given rays showing whether and how a $(6n+2)$-ray model can be built. To be specific,
from (iii) to (xii), we  show that if the orthogonality graph (see FIG.\ref{CCli}) is given, then there should be a constraint between the product of two outer rays  and the product of two  inner ones (notice that two inner rays $\textbf{e},\textbf{f}$ in the case of order $n=k$ can also be considered as two outer rays $\textbf{a},\textbf{b}$ in the case of order $n=k-1$). This constraint will help us to get the minimal number of the auxiliary complete orthonormal bases. And in (xiii), given two (outer) rays, we show how to recursively choose the inner rays to build a $(6n+2)$-ray model.

\section{Appendix B. Proof of Inequality Eq.(2)}

\begin{figure}[h]
\includegraphics[scale=1]{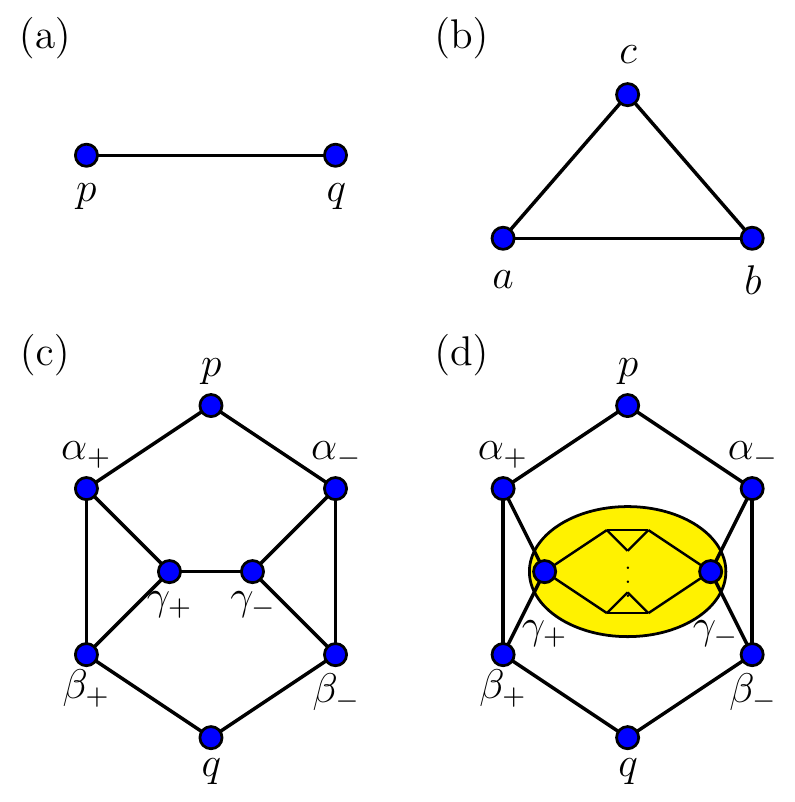} \caption{\label{app-sdinq}  Each graph gives the related  KS value assignment inequality, where the vertices represent the rays and the solid lines  indicate the orthogonality for them. (a)Two orthogonal rays. (b)A complete orthonormal base. (c) The Clifton's 8-ray model. (d) The $(6n+2)$-ray model.  }
\end{figure}

We can derive the corresponding inequalities by algebraic proofs. We denote by each index of the vertex in FIG.\ref{app-sdinq} the related observable. Each observable can only be assigned to value 0 or 1 due to the NCHVT, e.g., $p,q\in\mZ_2\equiv\{0,1\}$ for FIG.\ref{app-sdinq}-(a), etc. For simplicity, hereafter we may use the same notation to represent an observable and its the expectation value, e.g., $p$ and $\langle p\rangle$. One can easily distinguish them from the context.

(a) The relevant KS value assignment inequality for FIG.\ref{app-sdinq}-(a)  can be written as
\begin{align*}
    p+q-pq=1-\bar{p}\bar{q}\leq1,
\end{align*}
where $\bar{p}=1-p,\bar{q}=1-q$.

(b)The KS value assignment inequality for the graph in FIG.\ref{app-sdinq}-(b) is
\begin{align*}
    \bigtriangleup-\Gamma_{\bigtriangleup}\equiv a+b+c-ab-ac-bc=1-abc-\bar{a}\bar{b}\bar{c}\leq1,
\end{align*}
where $\bar{a}=1-a,\bar{b}=1-b$ and $\bar{c}=1-c$.

(c)The KS value assignment inequality for FIG.\ref{app-sdinq}-(c) can be summarized as
\begin{align*}
    \langle \mB\rangle=&p+q+\bigtriangleup_++\bigtriangleup_--p(\alpha_++\alpha_-)-q(\beta_++\beta_-)\\
    &-\Gamma_{\bigtriangleup_+}-\Gamma_{\bigtriangleup_-}-\gamma_+\gamma_-\\
    \leq&3,
\end{align*}
where ${\bigtriangleup_{\pm}}=\alpha_{\pm}+\beta_{\pm}+\gamma_{\pm}$, $\Gamma_{\bigtriangleup_{\pm}}=\alpha_{\pm}\beta_{\pm}+\alpha_{\pm}\gamma_{\pm}+\beta_{\pm}\gamma_{\pm}$.

{\it Proof.---} (i) For $p+q\leq1$, from the results of (b), we have ${\bigtriangleup_{\pm}}-\Gamma_{\bigtriangleup_{\pm}}\leq1$, thus the proof is straightforward.

(ii) For $p+q=2~(p=q=1)$, we have
\begin{align*}
    \langle \mB\rangle=&2+(\alpha_++\alpha_-+\beta_++\beta_-+\gamma_++\gamma_-)\\
    &-(\alpha_++\alpha_-)-(\beta_++\beta_-)
    -\Gamma_{\bigtriangleup_+}-\Gamma_{\bigtriangleup_-}-\gamma_+\gamma_-\\
    =&2+(\gamma_++\gamma_--\gamma_+\gamma_-)-\Gamma_{\bigtriangleup_+}-\Gamma_{\bigtriangleup_-}\\
    \leq&3-\Gamma_{\bigtriangleup_+}-\Gamma_{\bigtriangleup_-}\\
    \leq&3.
\end{align*}
Here we have used the elementary inequality from (a), i.e., $\gamma_++\gamma_--\gamma_+\gamma_-\leq1$. \hfill$\sharp$

(d)The KS value assignment inequality for the $(6n+2)$-ray model in FIG.\ref{app-sdinq}-(d) is given as
\begin{align*}
    \langle \mB_n\rangle=&\sum_{i\in \mV}P_i-\sum_{i<j,(i,j)\in \mE}P_iP_j\\
     =&p+q+\bigtriangleup_++\bigtriangleup_--\Gamma_{\bigtriangleup_+}-\Gamma_{\bigtriangleup_-}-(\gamma_++\gamma_-)\\
     &-p(\alpha_++\alpha_-)-q(\beta_++\beta_-)
    +\langle \mB_{n-1}\rangle\\
    \leq&1+2n,
\end{align*}
where $\mV,\mE$ denote the vertex set and the edge set for the graph representation of the $(6n+2)$-ray model, $P_i$ and $(i,j)$ are the $i$-th ray (or vertex) and the edge between the $i$-th ray and the $j$-th ray, respectively. The recursive term $\langle \mB_{n-1}\rangle$ is due to the contribution of the yellow zone in FIG.\ref{app-sdinq}-(d).

{\it Proof.---} Notice that (c) is the case for $n=1$. Assume that the statement holds for $n-1$, i.e., $\langle \mB_{n-1}\rangle\leq1+2(n-1)$. Then we must show that it holds for $\langle \mB_n\rangle$.

(i) For $p+q=0~(p=q=0)$, using the results of (b), we can get
\begin{align*}
  \langle \mB_n\rangle\leq2+\langle \mB_{n-1}\rangle\leq2+1+2(n-1)\leq1+2n.
\end{align*}

(ii) For $p+q=1$, if we relabel all the complete orthonormal bases (triangles in the graph) from $1$ to $2n$ in FIG.\ref{app-sdinq}-(d), we can get
\begin{align*}
  \langle \mB_n\rangle=&1+\sum_{k=1}^{2n}(\bigtriangleup_k-\Gamma_{\bigtriangleup_k})
  -\sum_{k,l=1(k<l)}^{2n}\Gamma_{kl}\\
  &-p(\alpha_++\alpha_-)-q(\beta_++\beta_-)\\
  \leq&1+\sum_{k=1}^{2n}(\bigtriangleup_k-\Gamma_{\bigtriangleup_k})\\
  \leq&1+2n,
\end{align*}
where $\Gamma_{kl}$ is the product of the two rays(vertices) from the $k$-th and the $l$-th triangles respectively if there is a link between them, and zero otherwise. Here we also have used the results from (b).

(iii)For $p+q=2~(p=q=1)$,
\begin{align*}
      \langle\mB_n\rangle=&2+\bigtriangleup_++\bigtriangleup_--\Gamma_{\bigtriangleup_+}-\Gamma_{\bigtriangleup_-}\\
     &-(\alpha_++\alpha_-)-(\beta_++\beta_-)-(\gamma_++\gamma_-)
    +\langle \mB_{n-1}\rangle\\
    =&2-\Gamma_{\bigtriangleup_+}-\Gamma_{\bigtriangleup_-}+\langle \mB_{n-1}\rangle\\
    \leq&2+\langle \mB_{n-1}\rangle\\
    \leq&1+2n.
\end{align*}

From (i),(ii) and (iii), the statement is true for $\langle \mB_n\rangle$. \hfill$\sharp$

Next we discuss the quantum violation for this $(6n+2)$-ray model.

To be consist with the notations of Lemma in the main text. We denote by $|\psi\rangle\langle\psi|=p$ and $|\phi\rangle\langle\phi|=q$.
Then $\mB_n=K+C(\psi,\phi)$, where $K=|\psi\rangle\langle\psi|+|\phi\rangle\langle\phi|$. We can always choose a proper basis$\{|\psi\rangle,|\psi^{\perp}\rangle,|\psi^{\perp,\prime}\rangle\}$ such that $\langle\psi^{\perp,\prime}|\phi\rangle=0$, then
$|\phi\rangle=|\psi\rangle\langle\psi|\phi\rangle+|\psi^{\perp}\rangle\langle\psi^{\perp}|\phi\rangle$. We have
\begin{align*}
    K=\left(
        \begin{array}{ccc}
          1+|\langle\psi|\phi\rangle|^2 & \langle\psi|\phi\rangle\langle\phi|\psi^{\perp}\rangle & 0 \\
          \langle\psi^{\perp}|\phi\rangle\langle\phi|\psi\rangle & |\langle\psi^{\perp}|\phi\rangle|^2& 0 \\
          0 & 0 & 0 \\
        \end{array}
      \right).
\end{align*}
The maximal eigenvalue for $K$ is $\lambda_K^{\max}=1+|\langle\psi|\phi\rangle|$. This can be derived from the constraint $|\langle\psi|\phi\rangle|^2+|\langle\psi^{\perp}|\phi\rangle|^2=1$. Notice that the number of the auxiliary complete orthonormal bases is $2n$. Therefore, $\langle C(\psi,\phi)\rangle_q=2n$. Then we have $\langle K+C(\psi,\phi)\rangle_q\leq2n+1+|\langle\psi|\phi\rangle|$.

\section{Appendix C.  Proof of Inequality Eq.(5)}

\begin{figure}
\includegraphics[scale=0.7]{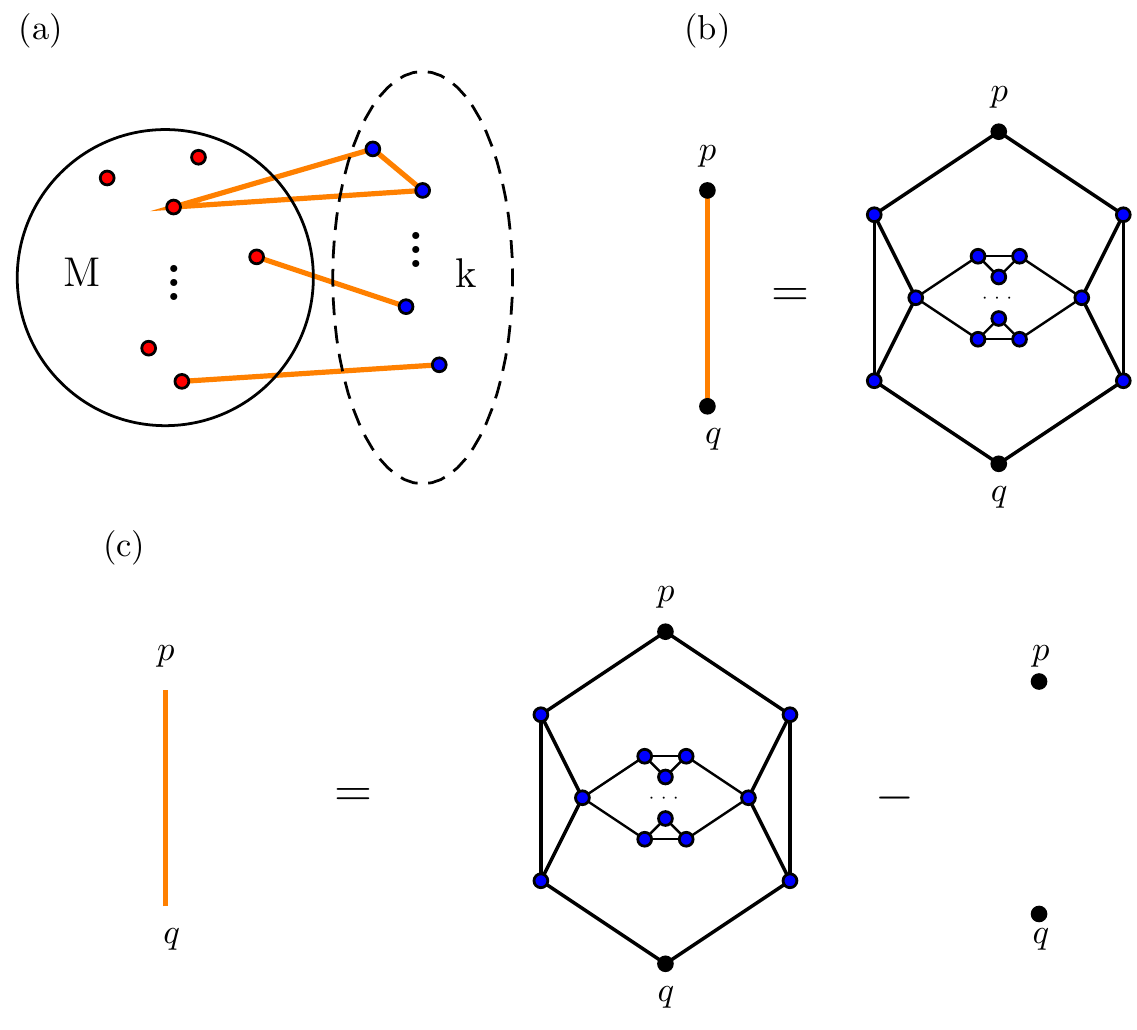} \caption{\label{app-siinq} (a) $M+k$ rays from an $n$-th order FKRS form a subset $\mK$, where each colored vertex represents a ray and each orange line represents a hyper-edge hereafter. Assume $M$ rays in the circle can form such a set $\mQ$ that for any two rays $|\psi_i\rangle$ and $|\psi_j\rangle$ in $\mQ$, they satisfy $|\langle\psi_i|{\psi}_j\rangle|>\delta_n=\frac{n}{n+2}$, where $M$ has been defined in the body of the Letter. (b)-(c)Graphical representation for a hyper-edge. Each black dot can be considered as a red or blue dot.}
\end{figure}

Given $n>0$ ($n\in\mZ$), for any two rays $|\psi_k\rangle$ and $|\psi_l\rangle$, if they satisfy $|\langle\psi_k|{\psi}_l\rangle|\in(\delta_{n-1},\delta_{n}]$, we can build a $(6n+2)$-ray model as shown in FIG.\ref{app-sdinq}-(d), where $p=|\psi_k\rangle\langle\psi_k|$ and $q=|\psi_l\rangle\langle\psi_l|$. Then an $n$-weighted {\it{hyper-edge}} linking these two rays (vertices) can be defined as all the complete orthonormal bases from this $(6n+2)$-ray model, see FIG.\ref{app-siinq}-(b)(c), where each orange line represents a hyper-edge (distinguished from ordinary definition for the edge of a graph).
This equivalence can be considered as the {\it hyper-graph} for the rays. Then the vertex set and the hyper-edge can be denoted by $V$ and $E$ respectively. To each hyper-graph $G$ we associate with the following NC observable
\begin{equation}
 G=\sum_{i\in V}P_i+\sum_{(i,j)\in E}C(P_i,P_j)
\end{equation}
where $C(P_i,P_j)$ is defined by Eq.(2) in main text and vanishes if $|\langle \psi_i|\psi_j\rangle|>\delta_n$ and involves $2n_{ij}$ complete orthonormal bases where
$$n_{ij}=\left\lceil\frac{2|\langle \psi_i|\psi_j\rangle|}{1-|\langle \psi_i|\psi_j\rangle|}\right\rceil.$$
As a result $C(P_i,P_j)\le 2n_{ij}$ and if there are at most $M$ projections in $V$ is assigned to value 1 then we have automatically $\langle G\rangle\le M+N$ with $N=2\sum_{i,j\in E}n_{ij}$ being the total number of complete orthonormal bases. Thus in what follows we always assume there are at least $M+1$ rays in $V$ that are assigned to value 1.

A {\it subgraph} $G^{\prime}$ of a graph $G$ is also a graph with a vertex set $V^{\prime}$ given by a subset of the vertex set for $G$
and an edge set specified by all the edges based on the vertex subset $V^{\prime}$ in $G$.
This definition can also be generalized to a hyper-graph, but for simplicity we still call such a corresponding sub-structure from a hyper-graph a ``subgraph" rather than a ``sub-hyper-graph".

We shall denote by $G_i$ the subgraph obtained by removing the vertex $i$ and all connecting edges and it holds
\begin{equation}
\langle G\rangle=\langle G_i\rangle+\langle P_i\rangle+\sum_{j\in \mN_i}\langle  C(P_i,P_j)\rangle
\end{equation}
from which it follows
\begin{eqnarray}
|V|\langle G\rangle&=&\sum_{i\in V}\langle G_i\rangle+\sum_{i\in V}\langle P_i\rangle+\sum_{i,j\in V}\langle  C(P_i,P_j)\rangle\nonumber\\
&=&\sum_{i\in V}\langle G_i\rangle-\sum_{i\in V}\langle P_i\rangle+2\langle G\rangle
\end{eqnarray}
or
\begin{eqnarray}\label{subg}
(|V|-2)\langle G\rangle&=&{\sum_{i\in V}\langle G_i\rangle-\sum_{i\in V}\langle P_i\rangle},
\end{eqnarray}
where $\mN_i$ is the neighborhood of the vertex $i$, i.e., if $j\in\mN_i$, then $(i,j)\in E$.

\begin{figure}
\includegraphics[scale=0.7]{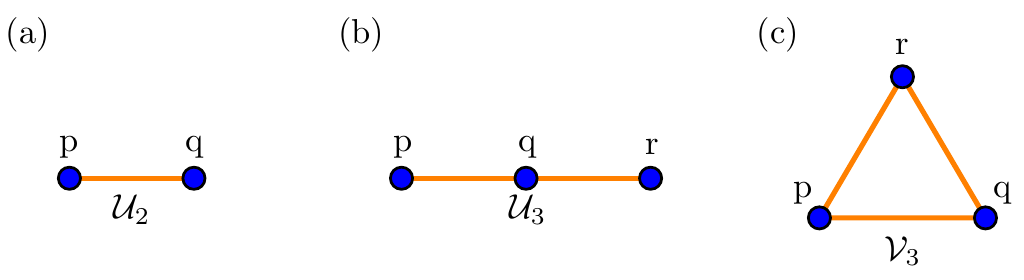} \caption{\label{app-openclosed} (a)-(c) Graphical representations for the $(6n+2)$-ray model $\mU_2$, the $[6(n_1+n_2)+3]$-ray model $\mU_3$ and the the $[6(n_1+n_2+n_3)+3]$-ray model $\mmV_3$. }
\end{figure}

Before proving our main result, we shall consider some simple examples of the hyper-graph in FIG.\ref{app-openclosed}, i.e., the two-vertex structure $\mU_2$, the three-vertex structures $\mU_3$ and $\mmV_3$.
The case for $\mU_2$  has been discussed in the previous section (FIG.\ref{app-sdinq}-(d)).  That is, we have $M=1$ and
\begin{align*}
  \langle \mU_2\rangle\leq1+2n,
\end{align*}
where we have used the new notation
$\langle \mU_2\rangle$ to replace  $\langle \mB\rangle$ in the above section.
For $\mU_3$ we have $M=2$ and the inequality is
\begin{align*}
  \langle \mU_3\rangle=&\langle \mB_{n_2}\rangle+p+\sum_{k=1}^{2n_1}(\bigtriangleup_k-\Gamma_{\bigtriangleup_k})
  -\sum_{k,l=1(k<l)}^{2n_1}\Gamma_{kl}\\
  &-p(\alpha_++\alpha_-)-q(\beta_++\beta_-)\\
    \leq&1+2n_2+1+2n_1\\
    \leq&2n_1+2n_2+2,
\end{align*}
where we have taken $q$, $r$ and the hyper-edge between them as a whole for consideration, i.e., the same structure as $\mU_2$. Besides, we use the similar discussion with paragraph (d) in the section of appendix B.

For $\mmV_3$ in FIG.\ref{app-openclosed}-(c), the KS value assignment inequality  can  be derived as follows,
\begin{align*}
  \langle \mmV_3\rangle=&p+q+r\\
  &+\sum_{k=1}^{2(n_1+n_2+n_3)}(\bigtriangleup_k-\Gamma_{\bigtriangleup_k})
  -\sum_{k,l=1(k<l)}^{2(n_1+n_2+n_3)}\Gamma_{kl}\\
  &-p(\alpha^{pq}_++\alpha^{pq}_-+\beta^{rp}_++\beta^{rp}_-)\\
  &-q(\alpha^{qr}_++\alpha^{qr}_-+\beta^{pq}_++\beta^{pq}_-)\\
  &-r(\alpha^{rp}_++\alpha^{rp}_-+\beta^{qr}_++\beta^{qr}_-),\\
\end{align*}
where $\alpha^{ab}_{\pm}$ and $\beta^{ab}_{\pm}~((a,b)\in\{(p,q),(q,r),(r,p)\})$  denote respectively the rays linking $a$ and $b$ from the set of rays corresponding to the hyper-edge $ab$. If $p+q+r\leq1$, it is clear that $\langle \mmV_3\rangle\leq2(n_1+n_2+n_3)+1$. For the case $p+q+r>1$, we can use another trick to derive the inequality,
\begin{align*}
  \langle \mmV_3\rangle=&\langle \mU_2(p,q)\rangle+\langle \mU_2(q,r)\rangle+\langle \mU_2(r,p)\rangle-(p+q+r)\cr
  \leq&2n_1+2n_2+2n_3+3-(p+q+r),
\end{align*}
where $\langle \mU_2(a,b)\rangle$ is a $\mU_2$ structure with vertices $a$ and $b$.  This indicates that $\mmV_3$ can  be  considered as $3$ $\mU_2$ structures minus the double-counted rays.
Thus, when $p+q+r>1$ ($p+q+r\geq2$),  we have
$\langle \mmV_3\rangle\leq2(n_1+n_2+n_3)+1$.
Finally, we can claim that the KS value assignment inequality for $\mmV_3$ is
\begin{align*}
  \langle \mmV_3\rangle\leq2(n_1+n_2+n_3)+1.
\end{align*}

\begin{figure}
\includegraphics[scale=0.6]{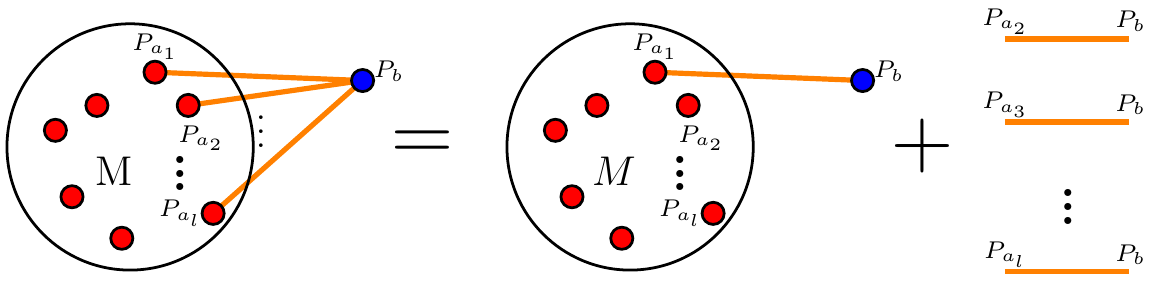} \caption{\label{M plus k proof} The $(M+k)$-vertex graphs with $k=|\mK\backslash\mQ|=1$}.
\end{figure}

This method used in $\mmV_3$ can be generalized to the case of $(M+k)$-vertex hyper-graph.  For the ray set $\mK=\{P_i\}_{i\in\mI}$ in FIG.\ref{app-siinq}-(a), where $\mI=\{1,2,...,M+k\}$,  we can draw at least $k$  hyper-edges (orange line) with weights equal to or less than $n$ by the definition of $M$ in the Letter. In what follows we shall prove by the method of induction that \begin{equation}\label{KSineq}
\langle G\rangle\le M+N_G.
\end{equation}
For fixed $M$ we shall at first show that it holds for all hyper-graphs $G_1$ on  $|V|=M+1$ vertices and then we shall prove that if it holds for all hyper-graphs $G_k$ on $|V|=M+k$ vertices  it is also true for all hyper-graphs $G_{k+1}$ on $|V|=M+k+1$ vertices.

In the case of $|V|=M+1$ the inequality can be proved with the help of FIG.\ref{M plus k proof}. We shall only have to consider the case in which all rays are assigned to value 1. Assume that $P_b$ does not belong to the maximal clique and thus there is at least one vertex $a_1$ that shares a hyper-edge with $b$. Let $P_{a_{1}},P_{a_{2}},...,P_{a_{l}}$ be the vertices adjacent to $P_b$. Then we have
 \begin{align*}
\langle G_1\rangle=&\sum_{i\neq a_{1},b}P_i+\langle\mU_2(P_{a_{1}},P_b)\rangle+\sum_{i=2}^{l}\langle C(P_{a_{i}},P_b)\rangle\\
&\leq (M+1-2)+1+2n_{a_1b}+2\sum_{i=2}^{l}n_{a_i,b}\\
&=M+N_1
\end{align*}
where the weight for the hyper-edges $P_bP_{a_{i}}$ is $n_i$.

Suppose that the Eq.(\ref{KSineq}) holds for all $G_k$, i.e., inequality $\langle G_k\rangle\le M+N_k$ holds for all hyper-graphs $G_k$ with $M+k$ vertices, and consider an arbitrary hyper-graph $G_{k+1}$ with $M+k+1$ vertices. We can suppose without loss of generality that
$\sum_{i\in V_{k+1}}P_i\geq M+1$. With the help of subgraph decomposition Eq.(\ref{subg}) we have
\begin{align*}
&(|V_{k+1}|-2)\langle G_{k+1}\rangle\\
=&\sum_{i\in V_{k+1}}\langle G_{k,i}\rangle-\sum_{i\in V_{k+1}}P_i\\
\leq& {|V_{k+1}|M}+\sum_{i\in V_{k+1}}N_{k,i}-(M+1)\\
\leq& {(|V_{k+1}|-2)M}+(|V_{k+1}|-2)N_{k+1}+M-1\\
=&(|V_{k+1}|-2)(M+N_{k+1})+M-1,
\end{align*}
Because in any possible value assignment $\langle G_{k+1}\rangle$ should be an integer and $|V_{k+1}|-2=M+k-1\ge M-1$, we obtain $\langle G_{k+1}\rangle\le M+N_{k+1}$.

\newpage


\begin{thebibliography}{99}
\bibitem{Bell}J. S. Bell, Physics \textbf{1}, 195 (1964).
\bibitem{CHSH}J. F. Clauser, M. A. Horne, A. Shimony, and R. A. Holt,   Phys. Rev. Lett. {\bf 23}, 880(1969).
\bibitem{Bell2}J. S. Bell, Rev. Mod. Phys. \textbf{38}, 447 (1966).
\bibitem{KS}S. Kochen and E.P. Specker, J. Math. Mech. \textbf{17}, 59 (1967).
\bibitem{mermin1}N.D. Mermin, Rev. Mod. Phys. \textbf{65}, 803 (1993).
\bibitem{Michler} M. Michler, H. Weinfurter, and M. \.{Z}ukowski, Phys. Rev. Lett. \textbf{84}, 5457
(2000).
\bibitem{Huang1} Y.F. Huang, C.F. Li, Y.S. Zhang, J.W. Pan, and G.C. Guo, Phys. Rev. Lett. \textbf{90}, 250401 (2003)
\bibitem{Kirchmair} G. Kirchmair, F. Z\"{a}hringer, R. Gerritsma, M. Kleinmann,
O. G\"{u}hne, A. Cabello, R. Blatt, and C. F. Roos, Nature
(London) \textbf{460}, 494 (2009).
\bibitem{Amselem} E. Amselem, M. R{\aa}dmark, M. Bourennane, and A.
Cabello, Phys. Rev. Lett. \textbf{103}, 160405 (2009).
\bibitem{Bartosik} H. Bartosik, J. Klepp, C. Schmitzer, S. Sponar, A. Cabello,
H. Rauch, and Y. Hasegawa, Phys. Rev. Lett. \textbf{103}, 040403
(2009).
\bibitem{Moussa} O. Moussa, C. A. Ryan, D. G. Cory, and R. Laflamme,
Phys. Rev. Lett. \textbf{104}, 160501 (2010).
\bibitem{Lapkiewicz} R. Lapkiewicz, P. Li, C. Schaeff, N. K. Langford, S.
Ramelow, M. Wie$\acute{s}$niak, and A. Zeilinger, Nature
(London) \textbf{474}, 490 (2011).
\bibitem{Amselem2} E. Amselem, L. E. Danielsen, A. J. L\'{o}pez-Tarrida, J. R.
Portillo, M. Bourennane, and A. Cabello, Phys. Rev. Lett. \textbf{108},
200405 (2012).
\bibitem{Zu} C. Zu, Y.-X.Wang, D.-L. Deng, X.-Y. Chang, K. Liu, P.-Y.
Hou, H.-X. Yang, and L.-M. Duan, Phys. Rev. Lett. \textbf{109}, 150401
(2012).
\bibitem{Vincenzo} V. D'Ambrosio, I. Herbauts, E. Amselem, E. Nagali,
M. Bourennane, F. Sciarrino, and A. Cabello,  Phys. Rev. X  \textbf{3}, 011012 (2013).
\bibitem{XiangZhang} X. Zhang, M. Um, J.H. Zhang, S. An, Y. Wang, D.-L. Deng,
C. Shen, L.-M. Duan, and K. Kim, Phys. Rev. Lett. \textbf{110}, 070401 (2013).
\bibitem{Huang2} Y.-F. Huang, M. Li, D.-Y. Cao, C. Zhang, Y.-S. Zhang, B.-H. Liu,
C.-F. Li, and G.-C. Guo, Phys. Rev. A \textbf{87}, 052133 (2013).
\bibitem{Peres} A. Peres, J. Phys. A \textbf{24}, L175 (1991).
\bibitem{Bub} J. Bub, Found. Phys. 26, 787 (1996).
\bibitem{Conway}J. H. Conway and S. Kochen, in Quantum [Un]speakables: From Bell to
Quantum Information, edited by R. A. Bertlmann and A. Zeilinger (Springer-Verlag, Berlin, 2002), p. 257.
\bibitem{yu-oh}S. Yu and C.H. Oh, Phys. Rev. Lett. \textbf{108}, 030402 (2012).
\bibitem{KCBS} A. A. Klyachko, M. A. Can, S. Binicio\v{g}lu, and A. S. Shumovsky, Phys. Rev. Lett. \textbf{101}, 020403
(2008).
\bibitem{Liang} Y.-C. Liang, R.W. Spekkens, and H. M. Wiseman, Phys.
Rep. \textbf{506}, 1 (2011).
\bibitem{Clifton} R. Clifton, Am. J. Phys. \textbf{61}, 443 (1993).
\bibitem{Simon} C. Simon, \v{C}. Brukner, and A. Zeilinger, Phys. Rev. Lett.
\textbf{86}, 4427 (2001).
\bibitem{Larsson} J. {\AA}. Larsson, Europhys. Lett. \textbf{58}, 799 (2002).
\bibitem{cabello2} A. Cabello, Phys. Rev. Lett. \textbf{101}, 210401 (2008).
\bibitem{cabello3} P. Badzi\c{a}g, I. Bengtsson, A. Cabello, and I. Pitowsky,
Phys. Rev. Lett. \textbf{103}, 050401 (2009).
\bibitem{TYO} W. Tang, S. Yu, and C. H. Oh, Phys. Rev. Lett. \textbf{110}, 100403 (2013).
\bibitem{Perespla} A. Peres, Phys. Lett. A 151, 107 (1990).
\bibitem{mermin1990} N.D. Mermin, Phys. Rev. Lett. {\bf 65}, 3373 (1990).
\bibitem{Peresbook} A. Peres, Quantum Theory: Concepts and Methods (Kluwer, Dordrecht, 1993).
\bibitem{Pusey} M.F. Pusey, Phys. Rev. Lett. \textbf{113}, 200401 (2014).
\bibitem{hardylike} A. Cabello, P. Badzi\c{a}g, M. Terra Cunha, and M. Bourennane, Phys. Rev. Lett. \textbf{111}, 180404 (2013).
\bibitem{NS condi} A. Cabello, M. Kleinmann, and C. Budroni, Phys. Lett. Lett. \textbf{114}, 250402(2015).

\bibitem{AEG} A. Cabello, J.M. Estebaranz, and G. Garc\'{\i}a-Alcaine, Phys. Lett. A \textbf{339}, 425 (2005).
\bibitem{PCs2}P. Lisonek, P. Badziag, J.R. Portillo, and A. Cabello, Phys. Rev. A \textbf{89}, 042101 (2014).
\bibitem{SIC21} I. Bengtsson, K. Blanchfield, and A. Cabello, Phys. Lett. A \textbf{376}, 374 (2012).
\bibitem{KP} M. Kernaghan, and A. Peres, Phys. Lett. A \textbf{198}, 1 (1995).

\bibitem{HD} S. Yu, and C.H. Oh, arXiv:1112.5513.

\end{thebibliography}
\end{document}